# Demonstration of multi-cavity optoelectronic oscillators based on multicore fibers

Sergi García, Javier Hervás and Ivana Gasulla
ITEAM Research Institute
Universitat Politècnica de València, Valencia, Spain
sergarc3@iteam.upv.es

*Abstract*— We report the first experimental demonstration of multi-cavity optoelectronic oscillators where the different cavities are hosted in a single multicore fiber. Different configurations are implemented on the same 20-m 7-core fiber link, exploiting both unbalanced dual-cavity operation (loop lengths are a multiple of a reference value) and multi-cavity Vernier operation (loop lengths are slightly different).

## I. Introduction

Multicore fibers (MCF) were originally envisioned to provide high-capacity digital communications over core and metro networks, [1]. Since they offer a compact medium for the propagation of parallel lightwave paths under the same environmental and mechanical conditions, they also find application in a wide range of areas including fiber-wireless access network distribution and Multiple Input Multiple Output antenna connectivity, multi-parameter fiber sensing and Microwave Photonics signal processing based on optical true time delay lines, [2-4].

One of the particularly attractive applications of Microwave Photonics is the generation of radiofrequency (RF) signals with a high spectral purity in a feedback-loop structure known as Optoelectronic Oscillator (OEO), [5-10]. Originally proposed by Yao and Maleki in a single-cavity configuration [5], OEOs can provide stable, tunable and low-linewidth RF generation along a broad RF range. High spectral-purity generation in single-cavity OEOs requires a long fiber loop, what generates a considerable number of oscillation modes separated by a small frequency period. This requires the incorporation of very selective RF filters to selects the oscillation mode for operation at a single frequency after photodetection. Multi-cavity or multi-loop OEOs were conceived to alleviate the narrowband requirement for the RF filter. In dual-cavity OEOs, a short cavity provides the required spectral separation between adjacent oscillating modes while the long cavity provides the required spectral purity [6]. This scheme can be generalized to an arbitrary number of cavities whose lengths are multiple of a given reference value, [7].

We can bring together the multiple cavities under the same fiber cladding by exploiting the spatial diversity inherent to MCFs. In [8], we proposed and theoretically evaluated the implementation of multi-cavity OEOs where the required cavities are provided by the different cores of the MCF. This approach allows for the implementation of architectures where the cavity lengths are either a multiple of a given reference value (unbalanced operation), or slightly different by exploiting the Vernier effect, [9].

In this paper, we report, for the first time to our knowledge, the experimental demonstration of different multi-cavity OEO structures implemented over the same homogeneous 7-core fiber. We demonstrate both unbalanced dual-cavity operation and multi-cavity Vernier operation.

## II. Dual-Cavity Unbalanced OEOs

As theoretically presented in [8], a dual-cavity unbalanced OEO operation can be implemented in an *N*-core MCF if $k_1$ cores ($k_1 < N/2$) are linked to form the short cavity while the remaining $N$-$k_1$ cores comprise the long cavity. Fig. 1 shows the experimental setup used for the demonstration of dual-cavity unbalanced operation in a 20-m 7-core fiber link where 3 different configurations were implemented: 1-core and 6-core cavities ($k_1 = 1$) corresponding to 20- and 120-m cavities; 2-core and 5-core cavities ($k_1 = 2$) corresponding to 40- and 100-m cavities; and 3-core and 4-core cavities ($k_1 = 3$) corresponding to 60- and 80-m cavities.

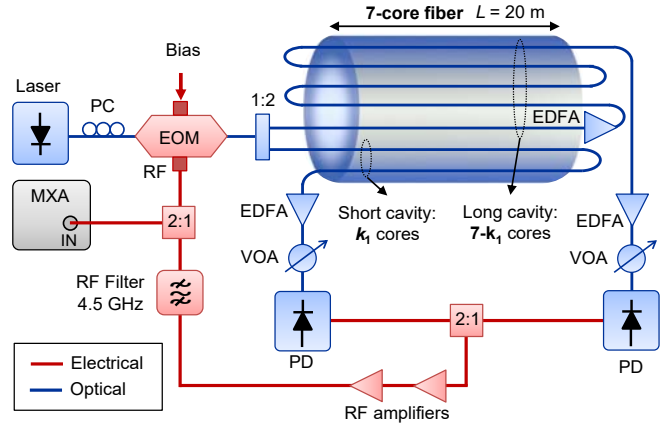

Fig. 1. Dual-cavity OEO based on the unbalanced cavity technique over a 20-meter 7-core homogeneous MCF. Three different configurations were demonstrated: 1-core and 6-core cavities ($k_1 = 1$), 2-core and 5-core cavities ($k_1 = 2$), 3-core and 4-core cavities ($k_1 = 3$). PC: polarization controller, EOM: electro-optic modulator, RF: radiofrequency, EDFA: Erbium-doped fiber amplifier, VOA: variable optical attenuator, PD: photodetector, MXA: Signal analyzer.

The MCF, which is provided by Fibercore, is characterized by a cladding diameter of 125 μm, a core separation of 35 μm, a core mode field diameter of 6.4 μm and a numerical aperture of 0.2. The fan-in/fan-out devices are provided by Optoscribe and have a maximum level of insertion loss of 2.5 dB in each way, including additional 1-dB losses due to the MCF splices. The measured intercore crosstalk, considering both the fan-in/fan-out device and the MCF, is lower than -50 dB. To compensate for the optical losses, mainly due to the fan-in/fan-out devices, optical amplification stages of 20 dB and 40 dB were included, respectively, in the short and long loops. After detection and the RF amplification stage, a tunable RF filter centered at 4.5 GHz with a bandwidth of 5 MHz allows for single-mode oscillation. A power splitter is set at the output of the RF filter to allow the continuous monitoring of the RF spectrum by using a signal analyzer (Agilent MXA signal analyzer N9020A, 20 Hz – 26.5 GHz). The cavity gain control is carried out in the optical domain by placing variable optical attenuators (VOA) before photodetection.

Figure 2 illustrates the measured RF oscillation spectra for the 3 dual-cavity unbalanced configurations over a 20-MHZ RF bandwidth, each one for 3 consecutive oscillating frequencies given by three different frequency tuning positions of the RF filter (blue, yellow and orange curves): (Upper) 1-core and 6-core cavities ($k_1 = 1$); (Middle) 2-core and 5-core cavities ($k_1 = 2$) and (Lower) 3-core and 4-core cavities ($k_1 = 3$). A spectrum periodicity of 4.8, 8.8, and 2.7 MHz, respectively for $k_1 = 1$, 2 and 3, is obtained. The fact that the configuration with the shortest short-cavity length (i.e., $k_1 = 1$) does not achieve the major Free Spectral Range (FSR) can be explained as follows. The FSR of the short and long cavities in isolation for $k_1 = 1$ are around 2.42 and 0.69 MHz, respectively, leading to mentioned FSR of 4.8 MHz when the dual-loop OEO is implemented. In contrast, the $k_1 = 2$ configuration leads to FSR values of 1.74 and 0.80 MHz for the isolated short and long cavities, respectively, so that their least common multiple determines the resulting 8.8-MHz FSR for the dual-loop OEO. Note that the short- and long-cavity delays (and thus their FSRs) are not only determined by the MCF length and the number of cores that form each cavity, but also by the inherent delays associated to the necessary components that form our setup, such as the internal length of the erbium doped fiber amplifiers (EDFAs) or the corresponding delay of the RF stage.

Both the computed and measured results of the phase noise spectra for a representative case ($k_1 = 2$) of the unbalanced OEO are shown in Fig. 3. As expected from [8], the measured phase noise results of all three configurations behave similarly. A phase noise around -80 dBc/Hz is achieved at a 10-KHz offset from the carrier, while it is downshifted below -120 dBc/Hz for frequency offsets above 1 MHz. The high level of phase noise close to the carrier can be explained by the fact that multiple amplification stages are used in our experimental setup. In particular, a double-stage EDFA amplification is required to build the long cavity, therefore, the current fluctuations of the pump current applied to the EDFAs may greatly contribute to the phase noise deterioration, [10].

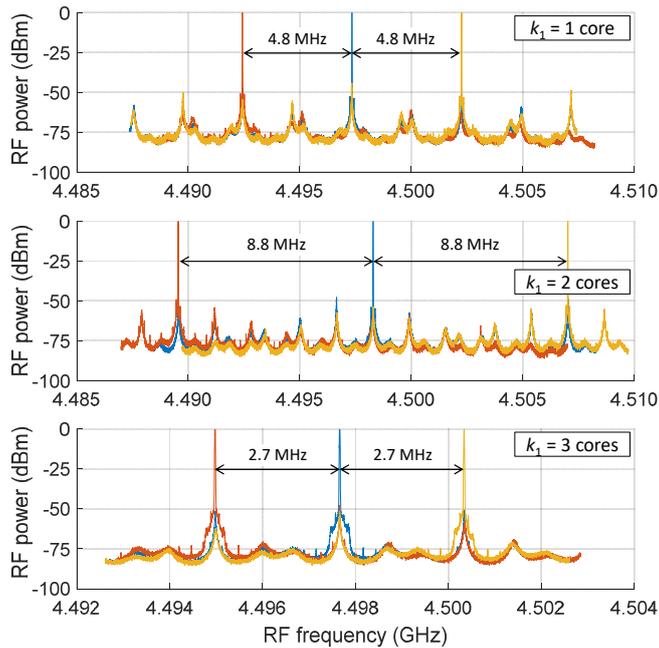

Fig. 2. Experimental oscillation spectra of a dual-cavity unbalanced OEO using a 20-meter 7-core homogeneous fiber for three different frequency tuning positions of the RF filter. (Upper) Configuration with 1-core and 6-core cavities ($k_1 = 1$). (Middle) Configuration with 2-core and 5-core cavities ($k_1 = 2$). (Lower) Configuration with 3-core and 4-core cavities ($k_1 = 3$).

For frequency offsets below 10 KHz, the phase noise performance is degraded by fluctuations that may occur in both fiber loops and the RF delay, as well as by the lack of proper temperature stabilization in the experimental setup. The peaks observed for frequency offsets above 800 KHz correspond to the minor oscillation modes present in both the short and long cavities, as we can also appreciate in Fig. 2.

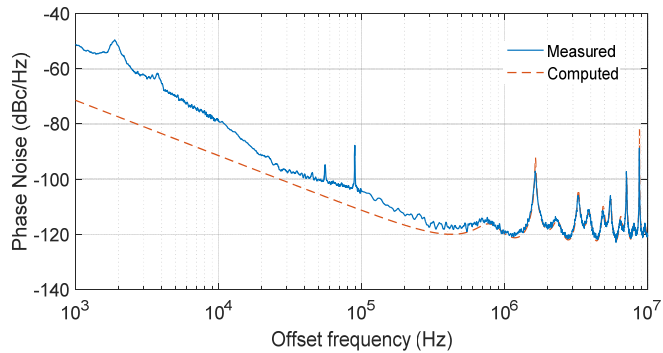

Fig. 3. Experimental and computed phase noise spectra of a dual-cavity unbalanced OEO using a 20-meter 7-core homogeneous fiber. Representative configuration with 2-core and 5-core cavities ($k_1 = 2$).

### III. MULTI-CAVITY VERNIER OEOs

A multi-cavity Vernier OEO can be implemented by using the $N$ cores of the MCF as the corresponding $N$ cavities that compose the oscillator. In the case of using a homogeneous MCF, the slightly different delays of the cavities can be obtained by adding an incremental length, $\Delta L$, to each cavity, which can be done compactly by properly designing the physical length of each fan-in fan-out device, [8].

Fig. 4 shows the experimental setup used to build a 3-cavity Vernier OEO in a 20-m homogeneous 7-core fiber, where three of its cores are used to perform the 3 cavities of the oscillator. For simplicity, we have used a 2-m long standard singlemode fiber to produce the required incremental delay between cavities and a variable delay line (VDL) to finely adjust the length difference between cavities. We have used the same MCF and fan-in/fan-out devices than in the unbalanced OEO, so that again an optical amplification stage of 20 dB is included in each loop to compensate their inherent insertion losses. Once the signals have been photodetected, coupled together and amplified, a 4.4 to 5.0 GHz tunable RF filter with a 30-MHz bandwidth is used to select the desired oscillation frequency.

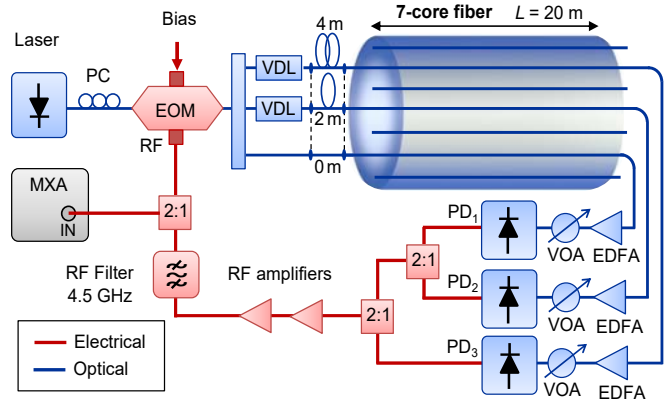

Fig. 4. Multi-cavity OEO based on the Vernier technique over a 7-core homogeneous MCF. Two different configurations were compared: 2-cavity and 3-cavity OEOs. PC: polarization controller, EOM: electro-optic modulator, RF: radiofrequency, VDL: variable delay line, EDFA: Erbium-doped fiber amplifier, VOA: variable optical attenuator, PD: photodetector, MXA: Signal analyzer.

We have measured the oscillation frequencies of 2-cavity and 3-cavity Vernier OEOs, as well as the RF spectrum corresponding to each one of the single cavities in isolation. The upper part of Fig. 5 illustrates the measured spectrum corresponding to each isolated cavity, where we see, as expected, that the OEO condition is not fulfil (with oscillation mode power levels lower than -70 dBm) as the open loop gain of each cavity is less than unity. With the help of the VDL placed at the first and second cavities, we have matched the oscillating modes of all three cavities at the frequency located near 4.499 GHz. The single-loop FSR is then 2.46, 2.40, and 2.34 MHz, respectively for isolated cavities 1, 2 and 3. This leads to a 2-cavity FSR of around 70 MHz and a global 3-cavity FSR above 600 MHz, so that a unique oscillating mode is sustained along the filter tuning range. The 30-MHz RF filter bandwidth is, by far, sufficiently narrow to allow single-mode oscillation for both the 2- and 3-cavitiy configurations.

To compare the spectrum of the oscillating modes of both 2-cavity and 3-cavity configurations, the lower part of Fig. 5 shows the measured RF spectrum for both OEOs around the 4.499-GHz oscillating mode when the central frequency of the RF filter is set at 4.5 GHz. With the help of the inset figure, we see clearly in a zoomed area within an offset frequency range of [-1 MHz, 1 MHz] from the carrier that both dual- and three-loop Vernier configurations provide very similar frequency responses.

Figure 6 shows the measured and computed phase noise spectra of both 2- and 3-cavity Vernier OEO configurations. As previously remarked for the case of the unbalanced dual-loop, the phase noise performance for frequency offsets below 10-20 KHz is highly degraded due to the effect of the EDFAs and the lack of a proper environmental isolation for the setup. For frequency offsets above 10 KHz, the computed and measured phase noise responses are in a good agreement and show clearly that both configurations provide a mostly equal phase noise behavior, as expected. This can be explained by the fact that Vernier configurations are based on having multiple cavities with quasi-identical delays, and since the phase noise basically depends on the longest cavity delay, their phase noise performance will be mostly identical.

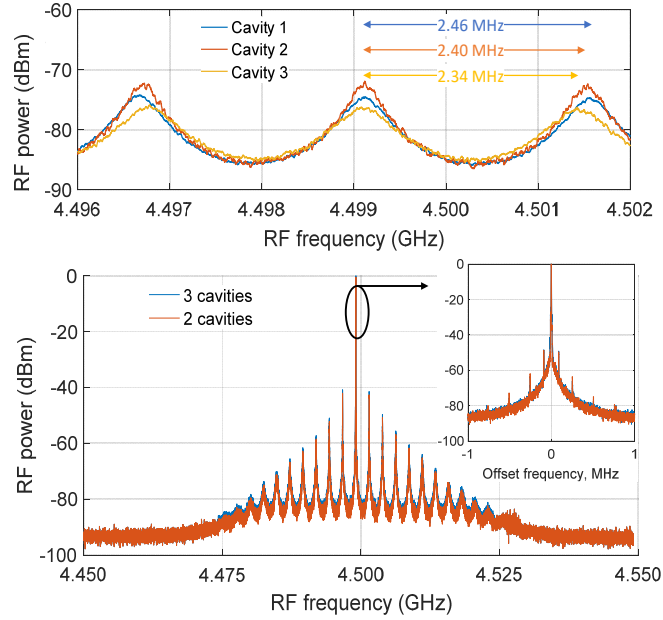

Fig. 5. Experimental oscillation spectra of a multi-cavity Vernier OEO using a 20-meter 7-core homogeneous fiber for (Upper) each of the three cavities in isolation, and (Lower) 2-cavity (red) and 3-cavity (blue) configurations. Inset: zoom area in the [-1 MHz, 1 MHz] offset frequency range.

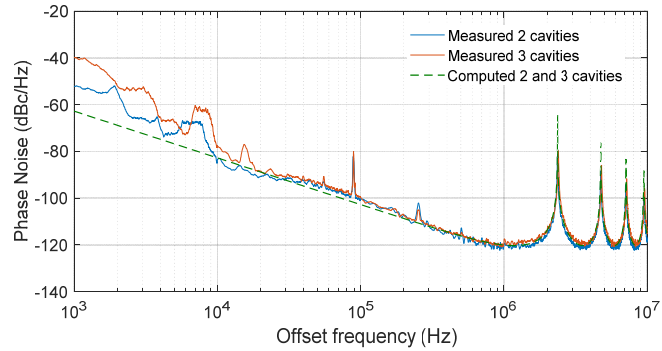

Fig. 6. Experimental and computed phase noise spectra of a multi-cavity Vernier OEO using a 20-meter 7-core homogeneous fiber for 2-cavity and 3-cavity configurations.

The measured phase noise at a 10-KHz offset is below -80 dBc/Hz and decreases as -10 dB/decade until reaching a lowest value of -120 dBc/Hz at 1-MHz frequency offset. We can see the similarity of these results compared to those shown in Fig. 3 for the unbalanced dual loop configurations. Although the phase noise values of both unbalanced and Vernier schemes are, in principle, not comparable to those reported for high-performance ultra-low phase noise OEOs [11], they can be significantly reduced by using low-residual phase noise components and by reducing the optical losses and thus suppressing the need for optical amplifiers. In the particular case of the Vernier OEO, we can furtherly contribute to diminishing the phase noise level by increasing the MCF length. In contrast to single-cavity OEOS, this will not alter the FSR of the OEO that will actually be preserved by the Vernier effect. In contrast, the proposed unbalanced OEO using a MCF could not benefit from a fiber length increase since it is configured using a relatively small number of fiber cores and, therefore, it is inherently linked to shorter cavity lengths to avoid the use of RF filters with extremely high selectivity. In that case, by increasing the number of fiber cores (in singlemode operation) up to 32, such as the latest reported dense-core singlemode MCFs [12], the length of the long cavity could be considerably increased as compared to the short one, improving as a consequence the phase noise performance.

## IV. CONCLUSIONS

To the best of our knowledge, we have presented here the first experimental demonstration of multi-cavity OEOs built upon multicore fibers. By exploiting the flexibility provided by the spatial diversity of the MCF, we demonstrated different OEO configurations in the same single 7-core fiber considering both highly unbalanced two-cavity operation as well as multi-cavity Vernier OEO operation, where moderate cavity lengths (20-120 m) are compatible with a high-spectral purity. Since all the cavities are hosted under the same cladding, the use of MCFs provides a fiber integrated hosting medium for enhancing the OEO relative stability against mechanical and environmental fluctuations.

ACKNOWLEDGMENTS

This research was supported by the ERC Consolidator Grant 724663, the Spanish Projects TEC2014-60378-C2-1-R and TEC2016-80150-R, the Spanish scholarships MECD FPU13/04675 for J. Hervás, MINECO BES-2015-073359 for S. García, and Spanish MINECO Ramón y Cajal RYC-2014-16247 for I. Gasulla. We thank Javier Madrigal for his work on the MCF fan-in/fan-out splices and Prof. Salvador Sales for his thoughtful discussions and recommendations.

REFERENCES

[1] D. J. Richardson, J. M. Fini, and L. E. Nelson, "Space division multiplexing in optical fibers," Nat. Photonics, vol. 7, pp. 354-362, 2013.
[2] J. Capmany et al., "Microwave photonic signal processing," J. Lightwave Technol., vol. 31, no. 4, pp. 571-586, 2013.
[3] I. Gasulla and J. Capmany, "Microwave photonics applications of multicore fibers," IEEE Photonics J., vol. 4, no. 3, pp. 877-887, 2012.
[4] S. Garcia et al., "Design of Heterogeneous Multicore Fibers as Sampled True-Time Delay Lines," Opt. Lett., vol. 40, no. 4, pp. 621-624, 2015.
[5] X.S. Yao and L. Maleki, "Optoelectronic microwave oscillator," J. Opt. Soc. Am. B, vol. 8, pp. 1725-1735, 1996.
[6] X.S. Yao and L. Maleki, "Multiloop optoelectronic oscillator," IEEE J. Quantum Electron., vol. 36, pp. 79-84, 2000.
[7] T. Bánky, B. Horváth and T. Berceli, "Optimum configuration of multiloop optoelectronic oscillators," J. Opt. Soc. Am. B, vol. 23, pp. 1371-1380, 2006.
[8] S. García and I. Gasulla, "Multi-cavity optoelectronic oscillators using multicore fibers," Opt. Express, vol. 23, pp. 2403-2415, 2015.
[9] Z. Tang et al., "Tunable Optoelectronic Oscillator Based on a Polarization Modulator and a Chirped FBG," IEEE Photonics Technol. Lett., vol. 24, pp. 1487-1489, 2012.
[10] W. Li and J. Yao, "An optically tunable optoelectronic oscillator," J. Lightwave Technol., vol. 28, no. 18, pp. 2640-2645, 2010.
[11] O. Lelièvre et al., "Ultra-low phase noise 10 GHz dual loop optoelectronic oscillator," Proc. of MWP 2016, TuMP9.
[12] T. Mizuno et al., "32-core dense SDM unidirectional transmission of PDM-16QAM signals over 1600 km using crosstalk-managed single-mode heterogeneous multicore transmission line," Proc. of OFC 2016, Th5C.3.